\documentclass{article}
\usepackage{spconf,amssymb,amsmath,graphicx,times,url,bm}
\usepackage{color,cite,soul}
\usepackage{booktabs,tabularx}
\usepackage{comment}
\usepackage{hyperref}
\usepackage[linesnumbered,ruled,vlined]{algorithm2e}
\usepackage{balance}

\newcommand{\ReduceSpaceUnderFigure}{\vspace{0pt}}
\newcommand{\ReduceSpaceUnderTable}{\vspace{-3pt}}
\newcommand{\ReduceSpaceAboveSubsection}{\vspace{-4pt}}
\DeclareMathOperator{\diag}{diag}

\DeclareMathOperator{\Fix}{Fix}
\DeclareMathOperator{\Id}{Id}

\makeatletter
\newcommand\footnoteref[1]{\protected@xdef\@thefnmark{\ref{#1}}\@footnotemark}
\makeatother

\title{WaveFit: An Iterative and Non-autoregressive Neural Vocoder \\ based on Fixed-Point Iteration}

\name{\parbox{0.99\linewidth}{
\centering
Yuma~Koizumi$^1$,~
Kohei~Yatabe$^2$,~
Heiga~Zen$^1$,~
Michiel~Bacchiani$^1$
}}
\address{$^1$Google Research, Japan \qquad $^2$Tokyo University of Agriculture and Technology, Japan}
\begin{document}

\ninept
\maketitle
\begin{sloppy}

\begin{abstract}
Denoising diffusion probabilistic models (DDPMs) and generative adversarial networks (GANs) are popular generative models for neural vocoders.
The DDPMs and GANs can be characterized by the iterative denoising framework and adversarial training, respectively.
This study proposes a fast and high-quality neural vocoder called \textit{WaveFit}, which integrates the essence of GANs into a DDPM-like iterative framework based on fixed-point iteration.
WaveFit iteratively denoises an input signal, and trains a deep neural network (DNN) for minimizing an adversarial loss calculated from intermediate outputs at all iterations.
Subjective (side-by-side) listening tests showed no statistically significant differences in naturalness between human natural speech and those synthesized by WaveFit with five iterations.
Furthermore, the inference speed of WaveFit was more than 240 times faster than WaveRNN.
Audio demos are available at \url{google.github.io/df-conformer/wavefit/}.
\end{abstract}


\section{Introduction}
\label{sec:intro}

%%%%%%%%%%%%%%% This paper focuses on neural vocoder. %%%%%%%%%%%%%%%
Neural vocoders~\cite{sample_rnn,tamamori2017speaker,waveglow,waveflow} are artificial neural networks that generate a speech waveform given acoustic features. They are indispensable building blocks of recent applications of speech generation. For example, they are used as the backbone module in text-to-speech (TTS)~\cite{tacotron2,nat,parallel_tacotron,pngbert,fastspeech,fastspeech2}, 
voice conversion~\cite{vc_overview,wenchin_vc_2022},
speech-to-speech translation (S2ST)~\cite{translatotron,translatotron2,lee-etal-2022-direct},
speech enhancement (SE)~\cite{Maiti_waspaa_2019,Maiti_icassp_2020,Su_2020,Su_2021}, 
speech restoration~\cite{voice_filxer,saeki2021_IS},
and speech coding~\cite{wavenet_codec,wavenet_lossless_codec,lpcnet_codec,sound_stream_codec}.
Autoregressive (AR) models first revolutionized the quality of speech generation~\cite{wavenet,sample_rnn,wavernn,lpcnet}.
However, as they require a large number of sequential operations for generation, parallelizing the computation is not trivial thus their processing time is sometimes far longer than the duration of the output signals.

%%%%%%%%%%%%%%% Related works: non-autoregressive neural vocoders. %%%%%%%%%%%%%%%
To speed up the inference, non-AR models have gained a lot of attention thanks to their parallelization-friendly model architectures.
Early successful studies of non-AR models are those based on normalizing flows~\cite{ParallelWaveNet,waveglow,waveflow} which convert an input noise to a speech using stacked invertible deep neural networks (DNNs)~\cite{normalizing_flow}.
In the last few years, the approach using generative adversarial networks (GANs)~\cite{gan_goodfellow2014} is the most successful non-AR strategy~\cite{Donahue_2019,Kong_2020,melgan,parallel_wavegan,Yang_2021,gan_vocoder,univnet,kaneko2022istftnet,avocado_vocoder,bigvgan} where they are trained to generate speech waveforms indistinguishable from human natural speech by discriminator networks.
The latest member of the generative models for neural vocoders is the denoising diffusion probabilistic model (DDPM)~\cite{wavegrad,diffwave,Bddm022,priorgrad,specgrad,okamoto2021,Goel_2022,InferGrad2022}, which converts a random noise into a speech waveform by the iterative sampling process as illustrated in Fig.~\ref{fig:concept}~(a).
With hundreds of iterations, DDPMs can generate speech waveforms comparable to those of AR models~\cite{wavegrad,diffwave}.

\begin{figure}[t]
\vspace{-1pt}
  \centering
  \includegraphics[width=\linewidth,clip]{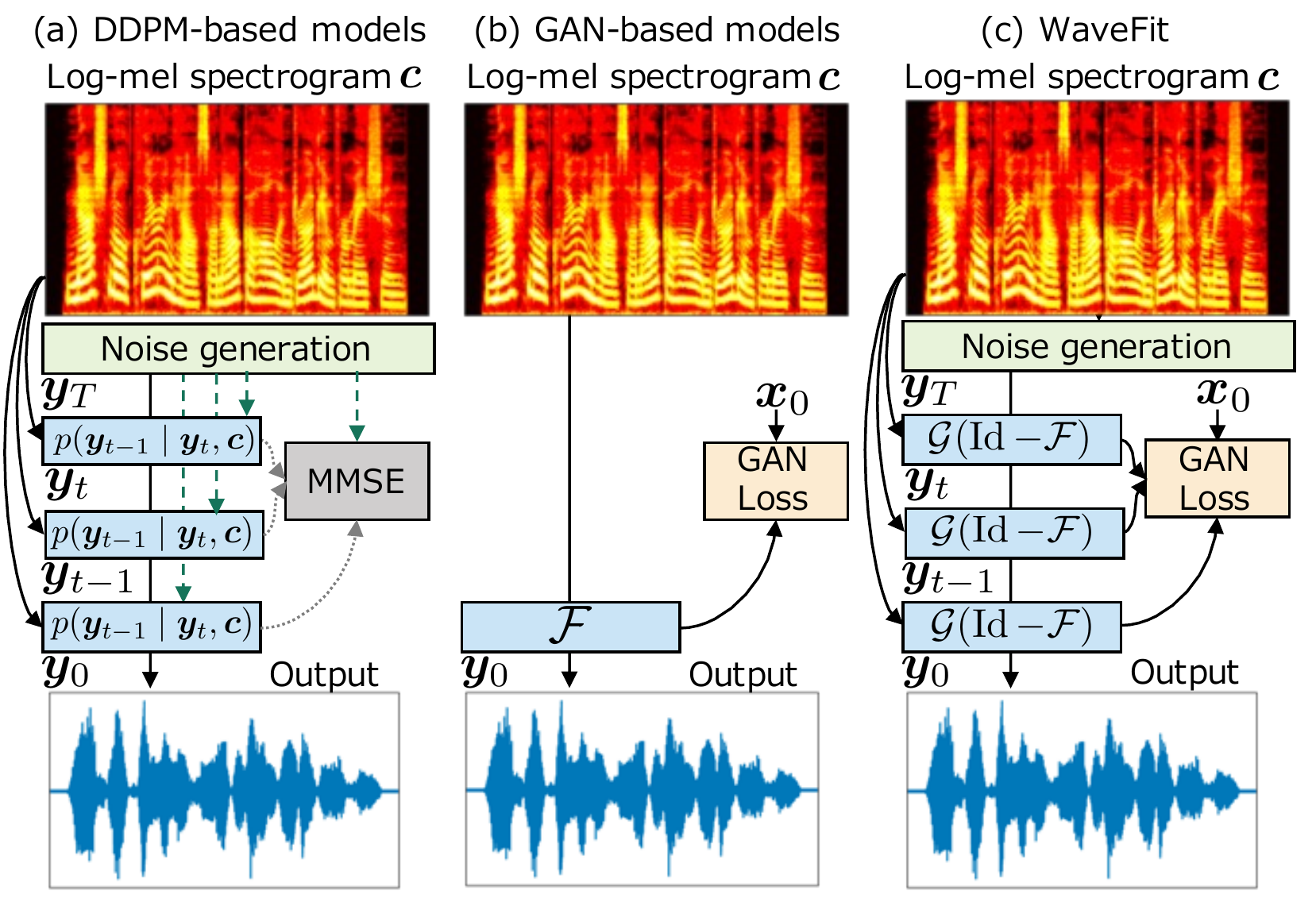} 
  \vspace{-20pt}
  \caption{Overview of (a) DDPM, (b) GAN-based model, and (c) proposed WaveFit.
  (a) DDPM is an iterative-style model, where sampling from the posterior is realized by adding noise to the denoised intermediate signals.
  (b) GAN-based models predict $\bm{y}_0$ by a non-iterative DNN $\mathcal{F}$ which is trained to minimize an adversarial loss calculated from $\bm{y}_0$ and the target speech $\bm{x}_0$. (c) Proposed WaveFit is an iterative-style model without adding noise at each iteration, and $\mathcal{F}$ is trained to minimize an adversarial loss calculated from all intermediate signals $\bm{y}_{T-1}, \ldots, \bm{y}_0 $, where $\Id$ and $\mathcal{G}$ denote the identity operator and a gain adjustment operator, respectively.}
  \label{fig:concept}
  \ReduceSpaceUnderFigure
\end{figure}

%%%%%%%%%%%%%%% Trade-off between the number of iterations and quality. %%%%%%%%%%%%%%%
%Although iterative-style waveform generation is a promising strategy, 
Since a DDPM-based neural vocoder iteratively refines speech waveform, there is a trade-off between its sound quality and computational cost~\cite{wavegrad}, i.e., tens of iterations are required to achieve high-fidelity speech waveform. To reduce the number of iterations while maintaining the quality, existing studies of DDPMs have investigated the inference noise schedule~\cite{Bddm022}, the use of adaptive prior~\cite{priorgrad,specgrad}, the network architecture~\cite{Goel_2022,okamoto2021}, and/or the training strategy~\cite{InferGrad2022}.
However, generating a speech waveform with quality comparable to human natural speech in a few iterations is still challenging.

%%%%%%%%%%%%%%% GAN + iterative-style is a promising strategy. %%%%%%%%%%%%%%%
Recent studies demonstrated that the essence of DDPMs and GANs can coexist~\cite{denoising_defusion_gans,diffgan_tts}. Denoising diffusion GANs~\cite{denoising_defusion_gans} use a generator to predict a clean sample from a diffused one and a discriminator is used to differentiate the diffused samples from the clean or predicted ones.
This strategy was applied to TTS, especially to predict a log-mel spectrogram given an input text~\cite{diffgan_tts}.
As DDPMs and GANs can be combined in several different ways, there will be a new combination which is able to achieve the high quality synthesis with a small number of iterations.

%%%%%%%%%%%%%%% We propose WaveFit %%%%%%%%%%%%%%%
This study proposes \textit{WaveFit}, an iterative-style non-AR neural vocoder, trained using a GAN-based loss as illustrated in Fig.~\ref{fig:concept}~(c).
It is inspired by the theory of \underline{\textbf{f}}ixed-point \underline{\textbf{it}}eration~\cite{FixedPoint_DataSci}. %, hence its name.
%The proposed model iteratively applies a deep neural network (DNN) as a denoising mapping to refine an input waveform so that the output is closer to the target speech.
The proposed model iteratively applies a DNN as a denoising mapping that removes noise components from an input signal so that the output becomes closer to the target speech.
We use a loss that combines a GAN-based~\cite{melgan} and a short-time Fourier transform (STFT)-based~\cite{parallel_wavegan} loss as this is insensitive to imperceptible phase differences.
By combining the loss for all iterations, the intermediate output signals are encouraged to approach the target speech along with the iterations.
Subjective listening tests showed that WaveFit can generate a speech waveform whose quality was better than conventional DDPM models.  The experiments also showed that the audio quality of synthetic speech by WaveFit with five iterations is comparable to those of WaveRNN~\cite{wavernn} and human natural speech.

%%%%%%%%%%%%%%%%%%%%%%%%%%%%%%%%%%%%%%%%%%%%%%%%%%%%%%%%%%%%%%%%%%%%%%%%
%%%%%%%%%%%%%%%%%%%%%%%%%%%%%%%%%%%%%%%%%%%%%%%%%%%%%%%%%%%%%%%%%%%%%%%%
%%%%%%%%%%%%%%%%%%%%%%%%%%% SECTION 2 %%%%%%%%%%%%%%%%%%%%%%%%%%%
%%%%%%%%%%%%%%% Review of non-AR neural vocoders %%%%%%%%%%%%%%%%
%%%%%%%%%%%%%%%%%%%%%%%%%%%%%%%%%%%%%%%%%%%%%%%%%%%%%%%%%%%%%%%%%%%%%%%%
%%%%%%%%%%%%%%%%%%%%%%%%%%%%%%%%%%%%%%%%%%%%%%%%%%%%%%%%%%%%%%%%%%%%%%%%
\section{Non-autoregressive neural vocoders}
\label{sec:non-auto}

%%%%%%%%%%%%%%% What non-autoregressive neural vocoder is. %%%%%%%%%%%%%%%
A \textit{neural vocoder} generates a speech waveform $\bm{y}_0 \in \mathbb{R}^D$ given a log-mel spectrogram $\bm{c} = (\bm{c}_1,...,\bm{c}_K) \in \mathbb{R}^{FK}$, where $\bm{c}_k \in \mathbb{R}^{F}$ is an $F$-point log-mel spectrum at $k$-th time frame, and $K$ is the number of time frames.
The goal is to develop a neural vocoder so as to generate $\bm{y}_0$ indistinguishable from the target speech $\bm{x}_0 \in \mathbb{R}^D$ with less computations.
This section briefly reviews two types of neural vocoders: DDPM-based and GAN-based ones.

%%%%%%%%%%%%%%% DDPM-based neural vocoders. %%%%%%%%%%%%%%%
\ReduceSpaceAboveSubsection
\subsection{DDPM-based neural vocoder}
\label{sec:ddpm_based}

%%%%%%%%%%%%%%% What is DDPM neural vocoder? %%%%%%%%%%%%%%%

A DDPM-based neural vocoder is a latent variable model of $\bm{x}_0$ as $q(\bm{x}_0 \mid \bm{c})$ based on a $T$-step Markov chain of $\bm{x}_{t} \in \mathbb{R}^D$ with learned Gaussian transitions, starting from $q(\bm{x}_T) = \mathcal{N}(\bm{0}, \bm{I})$, defined as
\begin{align}
    q(\bm{x}_0 \mid \bm{c}) = \int_{\mathbb{R}^{DT}} 
    q(\bm{x}_T) \prod_{t=1}^T q(\bm{x}_{t-1}  \mid \bm{x}_{t}, \bm{c}) \, \mathrm{d} \bm{x}_{1} \cdots \mathrm{d} \bm{x}_{T}.    
\end{align}
By modeling $q(\bm{x}_{t-1} \mid \bm{x}_{t}, \bm{c})$, $\bm{y}_0 \sim q(\bm{x}_0 \mid \bm{c})$ can be realized as a recursive sampling of $\bm{y}_{t-1}$ from $q(\bm{y}_{t-1} \mid \bm{y}_{t}, \bm{c})$.

%%%%%%%%%%%%%%% How to implement? %%%%%%%%%%%%%%%
In a DDPM-based neural vocoder, $\bm{x}_{t}$ is generated by the \textit{diffusion process} that gradually adds Gaussian noise to the waveform according to a noise schedule $\{\beta_1,...,\beta_T\}$ given by $p(\bm{x}_t \mid \bm{x}_{t-1}) = \mathcal{N}\left(\sqrt{1 - \beta_t}\bm{x}_{t-1}, \beta_t \bm{I} \right)$.
This formulation enables us to sample $\bm{x}_t$ at an arbitrary timestep $t$ in a closed form as 
$\bm{x}_t = \sqrt{\bar{\alpha}_{t}} \bm{x}_{0} + \sqrt{1 - \bar{\alpha}_{t}}\bm{\epsilon}$,
where $\alpha_t = 1 - \beta_t$, $\bar{\alpha}_{t} = \prod_{s=1}^t \alpha_s$, and $\bm{\epsilon} \sim \mathcal{N}(\bm{0}, \bm{I})$.
As proposed by Ho \textit{et al.}~\cite{Ho_2020}, DDPM-based neural vocoders use a DNN $\mathcal{F}$ with parameter $\theta$ for predicting $\bm{\epsilon}$ from $\bm{x}_t$ as $\hat{\bm{\epsilon}} = \mathcal{F}_{\theta}(\bm{x}_t, \bm{c}, \beta_t)$.
The DNN $\mathcal{F}$ can be trained by maximizing the evidence lower bound (ELBO), though most of DDPM-based neural vocoders use a simplified loss function which omits loss weights corresponding to iteration $t$;
\begin{align}
    \mathcal{L}^{\mbox{\tiny WG}} &= \left\lVert \bm{\epsilon} - \mathcal{F}_{\theta}(\bm{x}_t, \bm{c}, \beta_t) \right\rVert _2^2
    , \label{eq:wavegrad_loss}
\end{align}
where $\lVert \cdot \rVert_p$ denotes the $\ell_p$ norm.
% Note that WaveGrad~\cite{wavegrad} uses $\lVert \cdot \rVert_1$ instead of the mean-squared-error.
Then, if $\beta_t$ is small enough, $q(\bm{x}_{t-1} \mid \bm{x}_{t}, \bm{c})$ can be given by $\mathcal{N}(\bm{\mu}_{t}, \gamma_t \bm{I})$, and
the recursive sampling from $q(\bm{y}_{t-1} \mid \bm{y}_{t}, \bm{c})$ can be realized by iterating the following formula for $t=T,\dots,1$ as
\begin{align}
    \bm{y}_{t-1} &= \frac{1}{\sqrt{\alpha_t}} \left( \bm{y}_{t} - \frac{\beta_t}{\sqrt{1 - \bar{\alpha}_t}} \mathcal{F}_{\theta}(\bm{y}_{t}, \bm{c}, \beta_t) \right) + \gamma_t \bm{\epsilon}
    \label{eq:ddpm_synthesis}
\end{align}
where
$\gamma_t = \frac{1 - \bar{\alpha}_{t-1}}{1 - \bar{\alpha}_{t}}\beta_t$,
$\bm{y}_{T} \sim \mathcal{N}(\bm{0}, \bm{I})$ and $\gamma_1 = 0$.

%%%%%%%%%%%%%%% Few sentences to introduce other DDPM-based methods.  %%%%%%%%%%%%%%%
The first DDPM-based neural vocoders~\cite{wavegrad,diffwave} required over 200 iterations to match AR neural vocoders~\cite{wavenet,wavernn} in naturalness measured by mean opinion score (MOS). 
To reduce the number of iterations while maintaining the quality,
existing studies have investigated the use of noise prior distributions~\cite{priorgrad,specgrad} and/or better inference noise schedules~\cite{Bddm022}.

\ReduceSpaceAboveSubsection
\subsubsection{Prior adaptation from conditioning log-mel spectrogram}

%%%%%%%%%%%%%%% What is adaptive prior for DDPM? %%%%%%%%%%%%%%%
To reduce the number of iterations in inference, PriorGrad~\cite{priorgrad} and SpecGrad~\cite{specgrad} introduced an adaptive prior $\mathcal{N}(\bm{0}, \bm{\Sigma})$, where $\bm{\Sigma}$ is computed from $\bm{c}$.
The use of an adaptive prior decreases the lower bound of the ELBO, and accelerates both training and inference~\cite{priorgrad}.

%%%%%%%%%%%%%%% SpecGrad implementation %%%%%%%%%%%%%%%
SpecGrad~\cite{specgrad} uses the fact that $\bm{\Sigma}$ is positive semi-definite and that it can be decomposed as $\bm{\Sigma} = \bm{L} \bm{L}^{\top}$ where $\bm{L} \in \mathbb{R}^{D \times D}$ and ${}^{\top}$ is the transpose.
Then, sampling from $\mathcal{N}(\bm{0}, \bm{\Sigma})$ can be written as $\bm{\epsilon} = \bm{L} \tilde{\bm{\epsilon}}$ using $\tilde{\bm{\epsilon}} \sim \mathcal{N}(\bm{0}, \bm{I})$, and Eq.~\eqref{eq:wavegrad_loss} with an adaptive prior becomes
\begin{align}
    \mathcal{L}^{\mbox{\tiny SG}} &= \left\lVert \bm{L}^{-1} (\bm{\epsilon} - \mathcal{F}_{\theta}(\bm{x}_t, \bm{c}, \beta_t)) \right\rVert_2^2. \label{eq:specgrad_loss}
\end{align}
SpecGrad~\cite{specgrad} defines $\bm{L} = \bm{G}^{+}\bm{M}\bm{G}$ and approximates $\bm{L}^{-1} \approx \bm{G}^{+}\bm{M}^{-1}\bm{G}$.
Here 
$NK \times D$ matrix $\bm{G}$ represents the STFT, 
$\bm{M} = \diag[(m_{1,1}, \ldots, m_{N,K})] \in \mathbb{C}^{NK \times NK}$ is the diagonal matrix representing the filter coefficients for each $(n,k)$-th time-frequency (T-F) bin, 
and $\bm{G}^{+\!}$ is the matrix representation of the inverse STFT (iSTFT) using a dual window.
This means $\bm{L}$ and $\bm{L}^{-1}$ are implemented as time-varying filters and its approximated inverse filters in the T-F domain, respectively.
The T-F domain filter $\bm{M}$ is obtained by the spectral envelope calculated from $\bm{c}$ with minimum phase response.
The spectral envelope is obtained by applying the 24th order lifter to the power spectrogram calculated from $\bm{c}$.

\ReduceSpaceAboveSubsection
\subsubsection{InferGrad}

%%%%%%%%%%%%%%% Why InferGrad? %%%%%%%%%%%%%%%
In conventional DDPM-models, 
since the DNNs have been trained as a Gaussian denoiser using a simplified loss function as Eq.~\eqref{eq:wavegrad_loss}, there is no guarantee that the generated speech becomes close to the target speech. To solve this problem, InferGrad~\cite{InferGrad2022} synthesizes $\bm{y}_0$ from a random signal $\bm{\epsilon}$ via Eq.~\eqref{eq:ddpm_synthesis} in every training step, then additionally minimizes an infer loss $\mathcal{L}^{\mbox{\tiny IF}}$ which represents a gap between generated speech $\bm{y}_0$ and the target speech $\bm{x}_0$.
The loss function for InferGrad is given as
\begin{align}
    \mathcal{L}^{\mbox{\tiny IG}} = \mathcal{L}^{\mbox{\tiny WG}} + \lambda_{\mbox{\tiny IF}} \mathcal{L}^{\mbox{\tiny IF}}, \label{eq:infergrad_loss}
\end{align}
where $\lambda_{\mbox{\tiny IF}} > 0$ is a tunable weight for the infer loss.

%%%%%%%%%%%%%%% GAN-based neural vocoders. Non-iterative & non-autoregressive %%%%%%%%%%%%%%%
\ReduceSpaceAboveSubsection
\subsection{GAN-based neural vocoder}
\label{sec:gan_based}

%%%%%%%%%%%%%%% GAN-based models are non-iterative and non-AR models. %%%%%%%%%%%%%%%
Another popular approach for non-AR neural vocoders is to adopt adversarial training; a neural vocoder is trained to generate a speech waveform where discriminators cannot distinguish it from the target speech, and discriminators are trained to differentiate between target and generated speech.
In GAN-based models, a non-AR DNN $\mathcal{F}: \mathbb{R}^{FK} \to \mathbb{R}^D$ directly outputs $\bm{y}_{0}$ from $\bm{c}$ as $\bm{y}_{0} = \mathcal{F}_{\theta}(\bm{c})$.

%%%%%%%%%%%%%%% Recent GAN-based models use multi-scale cost functions. %%%%%%%%%%%%%%%
One main research topic with GAN-based models is to design loss functions.
Recent models often use multiple discriminators at multiple resolutions~\cite{melgan}. 
One of the pioneering work of using multiple discriminators is MelGAN~\cite{melgan} which proposed the multi-scale discriminator (MSD).
In addition, MelGAN uses a feature matching loss that minimizes the mean-absolute-error (MAE) between the discriminator feature maps of target and generated speech. The loss functions of the generator $\mathcal{L}^{\mbox{\tiny GAN}}_{\mbox{\tiny Gen}}$ and discriminator $\mathcal{L}^{\mbox{\tiny GAN}}_{\mbox{\tiny Dis}}$ of the GAN-based neural vocoder are given as followings:
\begin{align}
    \!\!\mathcal{L}^{\mbox{\tiny GAN}}_{\mbox{\tiny Gen}} \!&=\! \frac{1}{R_{\mbox{\tiny GAN}}}
    \sum _{r=1}^{R_{\mbox{\tiny GAN}}} - \mathcal{D}_r (\bm{y}_{0}) + \lambda_{\mbox{\tiny FM}} \mathcal{L}^{\mbox{\tiny FM}}_r(\bm{x}_{0}, \bm{y}_{0}) \label{eq:mel_gan_gen_loss}\\
    \!\!\mathcal{L}^{\mbox{\tiny GAN}}_{\mbox{\tiny Dis}} \!&=\! 
    \frac{1}{R_{\mbox{\tiny GAN}}}\!
    \sum _{r=1}^{R_{\mbox{\tiny GAN}}} \!
    \max(0, 1 \!-\! \mathcal{D}_r (\bm{x}_{0})) \!+ \max(0, 1 \!+\! \mathcal{D}_r (\bm{y}_{0})) \label{eq:mel_gan_dis_loss}
\end{align}
where $R_{\mbox{\tiny GAN}}$ is the number of discriminators and 
$\lambda_{\mbox{\tiny FM}} \ge 0$ is a tunable weight for $\mathcal{L}^{\mbox{\tiny FM}}$.
The $r$-th discriminator $\mathcal{D}_r: \mathbb{R}^D \to \mathbb{R}$
consists of $H$ sub-layers as
$\mathcal{D}_r = 
\mathcal{D}_r^{H} \circ \cdots \circ \mathcal{D}_r^{1}$ where 
$\mathcal{D}_r^{h}: \mathbb{R}^{D_{h-1, r}} \to \mathbb{R}^{D_{h, r}}$.
Then, the feature matching loss for the $r$-th discriminator is given by
\begin{align}
    \mathcal{L}^{\mbox{\tiny FM}}_r(\bm{x}_{0}, \bm{y}_{0}) = \frac{1}{H-1} \sum_{h=1}^{H-1} \frac{1}{D_{h, r}} \lVert \bm{d}_{x, 0}^{h} - \bm{d}_{y, 0}^{h} \rVert_1,
\end{align}
where $\bm{d}_{a, b}^{h}$ is the outputs of 
$\mathcal{D}_r^{h-1}(\bm{a}_b)$.

%%%%%%%%%%%%%%% Additionally uses multi-resolution STFT losses, proposed by Yamamoto-san. %%%%%%%%%%%%%%%
As an auxiliary loss function, a multi-resolution STFT loss is often used to stabilize the adversarial training process~\cite{parallel_wavegan}.
A popular multi-resolution STFT loss $\mathcal{L}^{\mbox{\tiny MR-STFT}}$ consists of the spectral convergence loss and the magnitude loss as~\cite{parallel_wavegan,Yang_2021,demucs}:
\begin{align}
    \mathcal{L}^{\mbox{\tiny MR-STFT}}(\bm{x}_{0}, \bm{y}_{0}) = 
     \frac{1}{R_{\mbox{\tiny STFT}}}
     \sum_{r=1}^{R_{\mbox{\tiny STFT}}}
    \mathcal{L}^{\mbox{\tiny Sc}}_r(\bm{x}_{0}, \bm{y}_{0}) + 
    \mathcal{L}^{\mbox{\tiny Mag}}_r(\bm{x}_{0}, \bm{y}_{0}),
\end{align}
where $R_{\mbox{\tiny STFT}}$ is the number of STFT configurations.
$\mathcal{L}^{\mbox{\tiny Sc}}_r$ and $\mathcal{L}^{\mbox{\tiny Mag}}_r$ correspond to the spectral convergence loss and the magnitude loss of the $r$-th STFT configuration as
$\mathcal{L}^{\mbox{\tiny Sc}}_r(\bm{x}_{0}, \bm{y}_{0}) = 
\lVert \bm{X}_{0, r} - \bm{Y}_{0, r} \rVert_2 / \lVert \bm{X}_{0, r} \rVert_2
$ and
$\mathcal{L}^{\mbox{\tiny Mag}}_r(\bm{x}_{0}, \bm{y}_{0}) = \frac{1}{N_r K_r} \lVert \ln(\bm{X}_{0, r}) - \ln(\bm{Y}_{0, r}) \rVert_1,$
where $N_r$ and $K_r$ are the numbers of frequency bins and time-frames of the $r$-th STFT configuration, respectively, and $\bm{X}_{0,r} \in \mathbb{R}^{N_r K_r}$ and $\bm{Y}_{0,r} \in \mathbb{R}^{N_r K_r}$ correspond to the amplitude spectrograms with the $r$-th STFT configuration of $\bm{x}_0$ and $\bm{y}_0$.

State-of-the-art GAN-based neural vocoders~\cite{Kong_2020} can achieve a quality nearly on a par with human natural speech. Recent studies showed that the essence of GANs can be incorporated into DDPMs~\cite{denoising_defusion_gans,diffgan_tts}. As DDPMs and GANs can be combined in several different ways, there will be a new combination which is able to achieve the high quality synthesis with a small number of iterations.

%%%%%%%%%%%%%%%%%%%%%%%%%%%%%%%%%%%%%%%%%%%%%%%%%%%%%%%%%%%%%%%%%%%%%%%%
%%%%%%%%%%%%%%%%%%%%%%%%%%%%%%%%%%%%%%%%%%%%%%%%%%%%%%%%%%%%%%%%%%%%%%%%
%%%%%%%%%%%%%%%%%%%%%%%%%%% SECTION 3 %%%%%%%%%%%%%%%%%%%%%%%%%%%
%%%%%%%%%%%%%%% Review of fixed-point iteration and RED %%%%%%%%%%%%%%%%
%%%%%%%%%%%%%%%%%%%%%%%%%%%%%%%%%%%%%%%%%%%%%%%%%%%%%%%%%%%%%%%%%%%%%%%%
%%%%%%%%%%%%%%%%%%%%%%%%%%%%%%%%%%%%%%%%%%%%%%%%%%%%%%%%%%%%%%%%%%%%%%%%
\section{Fixed-point iteration}
\label{sec:fixed_point_iteration}

Extensive contributions to data science have been made by fixed-point theory and algorithms~\cite{FixedPoint_DataSci}.
These ideas have recently been combined with DNN to design data-driven iterative algorithms~\cite{unplugged2018,ICML_PnP2019,PesquetMMO2021,degli,degli_jstsp,RED_PRO}.
Our proposed method is also inspired by them, and hence fixed-point theory is briefly reviewed in this section.

A fixed point of a mapping $\mathcal{T}$ is a point $\bm{\phi}$ that is unchanged by $\mathcal{T}$, i.e., $\mathcal{T}(\bm{\phi}) = \bm{\phi}$.
The set of all fixed points of $\mathcal{T}$ is denoted as
\begin{align}
    \Fix(\mathcal{T}) = \bigl\{\,\bm{\phi}\in\mathbb{R}^D\mid\mathcal{T}(\bm{\phi})=\bm{\phi}\,\bigr\}.
\end{align}
Let the mapping $\mathcal{T}$ be firmly quasi-nonexpansive \cite{ConvexAnalysisBook}, i.e., it satisfies
\begin{align}
    \left\|\mathcal{T}(\bm{\xi})-\bm{\phi}\right\|_2
    \leq
    \left\|\bm{\xi}-\bm{\phi}\right\|_2
    \label{eq:defQuasiNE}
\end{align}
for every $\bm{\xi}\in\mathbb{R}^D$ and every $\bm{\phi}\in\Fix(\mathcal{T})$ $(\neq\varnothing)$, and there exists a quasi-nonexpansive mapping $\mathcal{F}$ that satisfies $\mathcal{T} = \frac{1}{2}\Id + \frac{1}{2}\mathcal{F}$, where $\Id$ denotes the identity operator.
Then, for any initial point, the following fixed-point iteration converges to a fixed point of $\mathcal{T}$:%
\footnote{To be precise, $\Id-\mathcal{T}$ must be demiclosed at $\bm{0}$. Note that we showed the fixed-point iteration in a very limited form, Eq.~\eqref{eq:defFixIter}, because it is sufficient for explaining our motivation. For more general theory, see, e.g., \cite{RED_PRO,Moreau_Hybrid,ConvexAnalysisBook}.}
\begin{align}
    \bm{\xi}_{n+1} = \mathcal{T}(\bm{\xi}_{n}).
    \label{eq:defFixIter}
\end{align}
That is, by iterating Eq.~\eqref{eq:defFixIter} from an initial point $\bm{\xi}_0$, we can find a fixed point of $\mathcal{T}$ depending on the choice of $\bm{\xi}_0$.

An example of fixed-point iteration is the following proximal point algorithm which is a generalization of iterative refinement~\cite{bond_prox}:
\begin{align}
    \bm{\xi}_{n+1} = \mathrm{prox}_{\mathcal{L}}(\bm{\xi}_{n}),
    \label{eq:defPPA}
\end{align}
where $\mathrm{prox}_{\mathcal{L}}$ denotes the proximity operator of a loss function $\mathcal{L}$,
\begin{align}
    \mathrm{prox}_{\mathcal{L}}(\bm{\xi}) \in \arg\min_{\bm{\zeta}} \Bigl[\, \mathcal{L}(\bm{\zeta})+\frac{1}{2}\left\|\bm{\xi}-\bm{\zeta}\right\|_2^2 \,\Bigr].
    \label{eq:defProx}
\end{align}
If $\mathcal{L}$ is proper lower-semicontinuous convex, then $\mathrm{prox}_{\mathcal{L}}$ is firmly (quasi-) nonexpansive, and hence a sequence generated by the proximal point algorithm converges to a point in $\Fix(\mathrm{prox}_{\mathcal{L}})=\arg\min_{\bm{\zeta}}\mathcal{L}(\bm{\zeta})$, i.e., Eq.~\eqref{eq:defPPA} minimizes the loss function $\mathcal{L}$.
Note that Eq.~\eqref{eq:defProx} is a negative log-likelihood of maximum \textit{a posteriori} estimation based on the Gaussian observation model with a prior proportional to $\exp(-\mathcal{L}(\cdot))$.
That is, Eq.~\eqref{eq:defPPA} is an iterative Gaussian denoising algorithm like the DDPM-based methods, which motivates us to consider the fixed-point theory.

The important property of a firmly quasi-nonexpansive mapping is that it is attracting, i.e., equality in Eq.~\eqref{eq:defQuasiNE} never occurs: $\left\|\mathcal{T}(\bm{\xi})-\bm{\phi}\right\|_2<\left\|\bm{\xi}-\bm{\phi}\right\|_2$.
Hence, applying $\mathcal{T}$ always moves an input signal $\bm{\xi}$ closer to a fixed point $\bm{\phi}$.
In this paper, we consider a denoising mapping as $\mathcal{T}$ that removes noise from an input signal, and let us consider an ideal situation.
In this case, the fixed-point iteration in Eq.~\eqref{eq:defFixIter} is an iterative denoising algorithm, and the attracting property ensures that each iteration always refines the signal.
It converges to a clean signal $\bm{\phi}$ that does not contain any noise because a fixed point of the denoising mapping,  $\mathcal{T}(\bm{\phi}) = \bm{\phi}$, is a signal that cannot be denoised anymore, i.e., no noise is remained.
If we can construct such a denoising mapping specialized to speech signals, then iterating Eq.~\eqref{eq:defFixIter} from any signal (including random noise) gives a clean speech signal, which realizes a new principle of neural vocoders.

% -------------------------------------------------------------------------
\section{Proposed Method}
\label{sec:propose}

This section introduces the proposed iterative-style non-AR neural vocoder, \textit{WaveFit}.
Inspired by the success of a combination of the fixed-point theory and deep learning in image processing~\cite{RED_PRO}, we adopt a similar idea for speech generation.
As mentioned in the last paragraph in Sec.~\ref{sec:fixed_point_iteration}, the key idea is to construct a DNN as a denoising mapping satisfying Eq.~\eqref{eq:defQuasiNE}.
% That is, WaveFit iteratively refine a noise waveform to generate a speech waveform.
We propose a loss function which approximately imposes this property in the training.
Note that the notations from Sec.~\ref{sec:non-auto} (e.g., $\bm{x}_0$ and $\bm{y}_T$) are used in this section.

%%%%%%%%%%%%%%% Method overview %%%%%%%%%%%%%%%
\ReduceSpaceAboveSubsection
\subsection{Model overview}
\label{sec:wavefit}

The proposed model iteratively applies a denoising mapping to refine $\bm{y}_t$ so that $\bm{y}_{t-1}$ is closer to $\bm{x}_0$.
By iterating the following procedure $T$ times, WaveFit generates a speech signal $\bm{y}_0$:
\vspace{-2pt}
\begin{align}
    \bm{y}_{t-1} = \mathcal{G}\left( \bm{z}_{t}, \bm{c} \right),\qquad\quad
    \bm{z}_{t} = \bm{y}_{t} - \mathcal{F}_{\theta}(\bm{y}_{t}, \bm{c}, t)
\end{align}
\vspace{-2pt}where $\mathcal{F}_{\theta}: \mathbb{R}^{D} \to \mathbb{R}^{D}$ is a DNN trained to estimate noise components, $\bm{y}_T \sim \mathcal{N}(\bm{0}, \bm{\Sigma})$, and $\bm{\Sigma}$ is given by the initializer of SpecGrad~\cite{specgrad}.
$\mathcal{G}\left( \bm{z}, \bm{c} \right): \mathbb{R}^{D} \to \mathbb{R}^{D}$ is a gain adjustment operator that adjusts the signal power of $\bm{z}_{t}$ to that of the target signal defined by $\bm{c}$.
Specifically, 
the target power $P_c$ is calculated from the power spectrogram calculated from $\bm{c}$.
Then, the power of $\bm{z}_{t}$ is calculated as $P_z$, and the gain of $\bm{z}_{t}$ is adjusted as $\bm{y}_t = (P_c / (P_z + s))\,\bm{z}_t$ where $s = 10^{-8}$ is a scalar to avoid zero-division.

\ReduceSpaceAboveSubsection
\subsection{Loss function}

WaveFit can obtain clean speech from random noise $\bm{y}_T$ by fixed-point iteration if the mapping $\mathcal{G}(\Id-\mathcal{F}_{\theta})$ is a firmly quasi-nonexpansive mapping as described in Sec.~\ref{sec:fixed_point_iteration}.
Although it is difficult to guarantee a DNN-based function to be a firmly quasi-nonexpansive mapping in general, we design the loss function to approximately impose this property.

%%%%%%%%%%%%%%% Loss is calculated from all outputs. %%%%%%%%%%%%%%%
The most important property of a firmly quasi-nonexpansive mapping is that an output signal $\bm{y}_{t-1}$ is always closer to $\bm{x}_{0}$ than the input signal $\bm{y}_{t}$.
To impose this property on the denoising mapping, we combine loss values for all intermediate outputs $\bm{y}_{0},\bm{y}_{1},\ldots,\bm{y}_{T-1}$ as follows:
\begin{align}
    \mathcal{L}^{\mbox{\tiny WaveFit}} = \frac{1}{T} \sum_{t=0}^{T-1} \mathcal{L}^{\mbox{\tiny WF}}(\bm{x}_0, \bm{y}_t).
    \label{eq:lossSumL}
\end{align}

The loss function $\mathcal{L}^{\mbox{\tiny WF}}$ is designed based on the following two demands: (i) the output waveform must be a high-fidelity signal; and (ii) the loss function should be insensitive to imperceptible phase difference.
The reason for second demand is as follow: the fixed-points of a DNN corresponding to a conditioning log-mel spectrogram possibly include multiple waveforms, because there are countless waveforms corresponding to the conditioning log-mel spectrogram due to the difference in the initial phase.
Therefore, phase sensitive loss functions, such as the squared-error, are not suitable as $\mathcal{L}^{\mbox{\tiny WF}}$.
Thus, we use a loss that combines GAN-based and multi-resolution STFT loss functions as this is insensitive to imperceptible phase differences:
\begin{align}
    \mathcal{L}^{\mbox{\tiny WF}}(\bm{x}_0, \bm{y}_t) = \mathcal{L}^{\mbox{\tiny GAN}}_{\mbox{\tiny Gen}}(\bm{x}_{0}, \bm{y}_{t}) +
    \lambda_{\mbox{\tiny STFT\,}} \mathcal{L}^{\mbox{\tiny STFT}}(\bm{x}_{0}, \bm{y}_{t}),
    \label{eq:wavefit_loss}
\end{align}
where $\lambda_{\mbox{\tiny STFT}} \ge 0$ is a tunable weight.

%%%%%%%%%%%%%%% Loss: SEANet (MelGAN)-based loss + MR-STFT loss. %%%%%%%%%%%%%%%
As the GAN-based loss function $\mathcal{L}^{\mbox{\tiny GAN}}_{\mbox{\tiny Gen}}$, a combination of multi-resolution discriminators and feature-matching losses are adopted. We slightly modified Eq.~\eqref{eq:mel_gan_gen_loss} to use a hinge loss as in SEANet~\cite{seanet}:
\begin{align}
    \mathcal{L}^{\mbox{\tiny GAN\!}}_{\mbox{\tiny Gen}}(\bm{x}_{0}, \bm{y}_{t}) =\! 
    \sum _{r=1}^{R}\! \max(0, 1 - \mathcal{D}_r (\bm{y}_{t})) + \lambda_{\mbox{\tiny FM}} \mathcal{L}^{\mbox{\tiny FM\!}}_r(\bm{x}_{0}, \bm{y}_{t}).\!\!
\end{align}
For the discriminator loss function, $\mathcal{L}^{\mbox{\tiny GAN}}_{\mbox{\tiny Dis}}(\bm{x}_{0}, \bm{y}_{t})$ in Eq.~\eqref{eq:mel_gan_dis_loss} is calculated and averaged over all intermediate outputs $\bm{y}_{0},\ldots,\bm{y}_{T-1}$.

%%%%%%%%%%%%%%% Detail of MR-STFT loss. %%%%%%%%%%%%%%%
For $\mathcal{L}^{\mbox{\tiny STFT}}$, we use $\mathcal{L}^{\mbox{\tiny MR-STFT}}$ because several studies showed that this loss function can stabilize the adversarial training~\cite{parallel_wavegan,demucs,Yang_2021}.
In addition, we use MAE loss between amplitude mel-spectrograms of the target and generated speech as used in HiFi-GAN~\cite{Kong_2020} and VoiceFixer~\cite{voice_filxer}.
Thus, $\mathcal{L}^{\mbox{\tiny STFT}}$ is given by
\begin{align}
    \mathcal{L}^{\mbox{\tiny STFT}}(\bm{x}_{0}, \bm{y}_{t}) = 
    \mathcal{L}^{\mbox{\tiny MR-STFT}}(\bm{x}_{0}, \bm{y}_{t}) +
     \frac{1}{FK} \lVert \bm{X}_0^{\mbox{\tiny Mel}} - \bm{Y}_t^{\mbox{\tiny Mel}} \rVert_1,
\end{align}
where $\bm{X}_0^{\mbox{\tiny Mel}} \in \mathbb{R}^{F K}$ and $\bm{Y}_t^{\mbox{\tiny Mel}} \in \mathbb{R}^{F K}$ denotes the amplitude mel-spectrograms of $\bm{x}_0$ and $\bm{y}_t$, respectively.

\ReduceSpaceAboveSubsection
\subsection{Differences between WaveFit and DDPM-based vocoders}
\label{sec:ddpm_and_wavefit}

Here, we discuss some important differences between the conventional and proposed neural vocoders.
The conceptual difference is obvious; DDPM-based models are derived using probability theory, while WaveFit is inspired by fixed-point theory.
Since fixed-point theory is a key tool for deterministic analysis of optimization and adaptive filter algorithms \cite{SPMag_adaptLearn}, WaveFit can be considered as an optimization-based or adaptive-filter-based neural vocoder.

We propose WaveFit because of the following two observations.
First, addition of random noise at each iteration disturbs the direction that a DDPM-based vocoder should proceed.
The intermediate signals randomly changes their phase, which results in some artifact in the higher frequency range due to phase distortion.
Second, a DDPM-based vocoder without noise addition generates notable artifacts that are not random, for example, a sine-wave-like artifact.
This fact indicates that a trained mapping of a DDPM-based vocoder does not move an input signal toward the target speech signal; it only focuses on randomness. % removing random components in the input signal.
Therefore, DDPM-based approaches have a fundamental limitation on reducing the number of iterations.

In contrast, WaveFit denoises an intermediate signal without adding random noise.
Furthermore, the training strategy realized by Eq.~\eqref{eq:lossSumL} faces the direction of denoising at each iteration toward the target speech.
Hence, WaveFit can steadily improve the sound quality by the iterative denoising.
These properties of WaveFit allow us to reduce the number of iterations while maintaining sound quality.

Note that, although the computational cost of one iteration of our WaveFit model is almost identical to that of DDPM-based models, training WaveFit requires more computations than DDPM-based and GAN-based models.
This is because the loss function in Eq.~\eqref{eq:lossSumL} consists of a GAN-based loss function for all intermediate outputs.
Obviously, computing a GAN-based loss function, e.g. Eq.~\eqref{eq:mel_gan_gen_loss}, requires larger computational costs than the mean-squared-error used in DDPM-based models as in Eq~\eqref{eq:wavegrad_loss}.
In addition, WaveFit computes a GAN-based loss function $T$ times.
Designing a less expensive loss function that stably trains WaveFit models should be a future work.

\ReduceSpaceAboveSubsection
\subsection{Implementation}
\label{sec:implementation}

\vspace{2pt}
\noindent
\textbf{Network architecture:}
We use ``WaveGrad Base model~\cite{wavegrad}'' for $\mathcal{F}$
, which has 13.8M trainable parameters.
To compute the initial noise $\bm{y}_T \sim \mathcal{N}(\bm{0}, \bm{\Sigma})$, we follow the noise generation algorithm of SpecGrad~\cite{specgrad}.

For each $\mathcal{D}_r$, we use the same architecture of that of MelGAN~\cite{melgan}.
$R_{\mbox{\tiny GAN}} = 3$ structurally identical discriminators are applied to input audio at different resolutions (original, 2x down-sampled, and 4x down-sampled ones). 
Note that the number of logits in the output of $\mathcal{D}_r$ is more than one and proportional to the length of the input. Thus, the averages of Eqs.~\eqref{eq:mel_gan_gen_loss} and \eqref{eq:mel_gan_dis_loss} are used as loss functions for generator and discriminator, respectively.

%%%%%%%%%%%%%%% Hyper parameters %%%%%%%%%%%%%%%
\vspace{2pt}
\noindent
\textbf{Hyper parameters:} 
We assume that all input signals are up- or down-sampled to 24 kHz.
For $\bm{c}$, we use $F=128$-dimensional log-mel spectrograms, where the lower and upper frequency bound of triangular mel-filterbanks are 20 Hz and 12 kHz, respectively. We use the following STFT configurations for mel-spectrogram computation and the initial noise generation algorithm; 50 ms Hann window, 12.5 ms frame shift, and 2048-point FFT, respectively.

For $\mathcal{L}^{\mbox{\tiny MR-STFT}}$, we use $R_{\mbox{\tiny STFT}} = 3$ resolutions, as well as conventional studies~\cite{parallel_wavegan,Yang_2021,demucs}.
The Hann window size, frame shift, and FFT points of each resolution are  [360, 900, 1800], [80, 150, 300], and [512, 1024, 2048], respectively. For the MAE loss of amplitude mel-spectrograms, we extract a 128-dimensional mel-spectrogram with the second STFT configuration.

% -------------------------------------------------------------------------
\section{Experiment}
\label{sec:experiment}

%%%%%%%%%%%%%%% Overview of experiment %%%%%%%%%%%%%%%

We evaluated the performance of WaveFit via subjective listening experiments. 
In the following experiments, we call a WaveFit with $T$ iteration as ``WaveFit-$T$''.
We used SpecGrad~\cite{specgrad} and InferGrad~\cite{InferGrad2022} as baselines of the DDPM-based neural vocoders,
Multiband (MB)-MelGAN~\cite{Yang_2021} and HiFi-GAN $V$1~\cite{Kong_2020} as baselines of the GAN-based ones,
and WaveRNN~\cite{wavernn} as a baseline of the AR one.
Since the output quality of GAN-based models is highly affected by hyper-parameters, we used well-tuned open source implementations of MB-MelGAN and HiFi-GAN published in \cite{kanbayashi}.
Audio demos are available in our demo page.\footnote{\label{footnote:demo}
\href{https://google.github.io/df-conformer/wavefit/}
{\texttt{google.github.io/df-conformer/wavefit/}}
}

\ReduceSpaceAboveSubsection
\subsection{Experimental settings}
\label{sec:experiment_setup}

\vspace{2pt}
\noindent
\textbf{Datasets:} We trained the WaveFit, DDPM-based and AR-based baselines with a proprietary speech dataset which consisted of 184 hours of high quality US English speech spoken by 11 female and 10 male speakers at 24 kHz sampling. For subjective tests, we used 1,000 held out samples from the same proprietary speech dataset.

To compare WaveFit with GAN-based baselines, the LibriTTS dataset~\cite{libritts} was used.
We trained a WaveFit-5 model from the combination of the ``train-clean-100'' and ``train-clean-360'' subsets at 24 kHz sampling.
For subjective tests, we used randomly selected 1,000 samples from the ``test-clean-100'' subset.
Synthesized speech waveforms for the ``test-clean-100'' subset published at~\cite{kanbayashi} were used as synthetic speech samples for the GAN-based baselines.
The file-name list used in the listening test is available in our demo page.\footnoteref{footnote:demo}

\vspace{2pt}
\noindent
\textbf{Model and training setup:} 
We trained all models using 128 Google TPU v3 cores with a global batch size of 512.
To accelerate training, we randomly picked up 120 frames (1.5 seconds, $D=36,000$ samples) as input. We trained WaveFit-2 and WaveFit-3 for 500k steps, and WaveFit-5 for 250k steps with the optimizer setting same as that of WaveGrad~\cite{wavegrad}.
The details of the baseline models are described in Sec.~\ref{sec:comparison_model_details}.

For the proprietary dataset, weights of each loss were $\lambda_{\mbox{\tiny FM}} = 100$ and $\lambda_{\mbox{\tiny STFT}} = 1$ based on the hyper-parameters of SEANet~\cite{seanet}.
For the LibriTTS, based on MelGAN~\cite{melgan} and MB-MelGAN~\cite{Yang_2021} settings, we used $\lambda_{\mbox{\tiny FM}} = 10$ and $\lambda_{\mbox{\tiny STFT}} = 2.5$, and excluded the MAE loss between amplitude mel-spectrograms of the target and generated waveforms.

\vspace{2pt} 
\noindent
\textbf{Metrics:} 
To evaluate subjective quality, we rated speech naturalness through MOS and side-by-side (SxS) preference tests.
The scale of MOS was a 5-point scale (1:~Bad, 2:~Poor, 3:~Fair, 4:~Good, 5:~Excellent) with rating increments of 0.5, and that of SxS was a 7-point scale (-3 to 3).
Subjects were asked to rate the naturalness of each stimulus after listening to it. 
Test stimuli were randomly chosen and presented to subjects in isolation, i.e., each stimulus was evaluated by one subject.
Each subject was allowed to evaluate up to six stimuli.
The subjects were paid native English speakers in the United States.
They were requested to use headphones in a quiet room.

In addition, we measured real-time factor (RTF) on an NVIDIA V100 GPU. We generated 120k time-domain samples (5 seconds waveform) 20 times, and evaluated the average RTF with 95 \% confidence interval. 

\ReduceSpaceAboveSubsection
\subsection{Details of baseline models}
\label{sec:comparison_model_details}

\vspace{2pt}
\noindent
\textbf{WaveRNN~\cite{wavernn}:}
The model consisted of a single long short-term memory layer with 1,024 hidden units, 5 convolutional layers with 512 channels as the conditioning stack to process the mel-spectrogram features, and a 10-component mixture of logistic distributions as its output layer.
It had 18.2M trainable parameters.
We trained this model using the Adam optimizer~\cite{kingma2014adam} for 1M steps.
The learning rate was linearly increased to $10^{-4}$ in the first 100 steps then exponentially decayed to $10^{-6}$ from 200k to 400k steps.

\vspace{2pt}
\noindent
\textbf{DDPM-based models:}
For both models, we used the same network architecture, optimizer and training settings with WaveFit. 

For SpecGrad~\cite{specgrad}, we used WG-50 and WG-3 noise schedules for training and inference, respectively. 
We used the same setups as \cite{specgrad} for other settings except the removal of the generalized energy distance (GED) loss~\cite{ged_loss} as we observed no impact on the quality.

We tested InferGrad~\cite{InferGrad2022} with two, three and five iterations.
The noise generation process, loss function, and noise schedules were modified from the original paper~\cite{InferGrad2022} as we could not achieve the reasonable quality with the setup from the paper.
We used the noise generation process of SpecGrad~\cite{specgrad}.
For loss function, we used $\mathcal{L}^{\mbox{\tiny SG}}$ instead of $\mathcal{L}^{\mbox{\tiny WG}}$, and used the WaveFit loss $\mathcal{L}^{\mbox{\tiny WF}}(\bm{x}_0, \bm{y}_0)$ described in Eq.~\eqref{eq:wavefit_loss} as the infer loss $\mathcal{L}^{\mbox{\tiny IF}}$.
Furthermore, according to \cite{InferGrad2022}, the weight for $\lambda_{\mbox{\tiny IF}}$ was $0.1$.
We used the WG-50 noise schedule~\cite{specgrad} for training.
For inference with each iteration, we used the noise schedule with \textsc{[3.e-04, 9.e-01]}, \textsc{[3.e-04, 6.e-02, 9.e-01]}, and \textsc{[1.0e-04, 2.1e-03, 2.8e-02, 3.5e-01, 7.0e-01]}, respectively, because the output quality was better than the schedules used in the original paper~\cite{InferGrad2022}.
As described in \cite{InferGrad2022}, we initialized the InferGrad model from a pre-trained checkpoint of SpecGrad (1M steps) then finetuned it for additional 250k steps. The learning rate was $5.0 \times 10^{-5}$.

\vspace{2pt}
\noindent
\textbf{GAN-based models:}
We used ``checkpoint-1000000steps'' from ``libritts\_multi\_band\_melgan.v2'' for MB-MelGAN~\cite{Yang_2021}, and ``checkpoint-2500000steps'' from ``libritts\_hifigan.v1'' for HiFi-GAN~\cite{Kong_2020}, respectively. These samples were stored in Google Drive linked from \cite{kanbayashi} (June 17th, 2022, downloaded). 

\ReduceSpaceAboveSubsection
\subsection{Verification experiments for intermediate outputs}
\label{sec:verification}

We first verified whether the intermediate outputs of WaveFit-5 were approaching to the target speech or not.
We evaluated the spectral convergence $\mathcal{L}^{\mbox{\tiny Sc}}_r$ and the log-magnitude absolute error $\mathcal{L}^{\mbox{\tiny Mag}}_r$ for all intermediate outputs.
The number of STFT resolutions was $R_{\mbox{\tiny STFT}} = 3$.
For each resolution, we used a different STFT configuration from the loss function used in InferGrad and WaveFit: the Hann window size, frame shift, and FFT points of each resolution were  [240, 480, 1,200], [48, 120, 240], and [512, 1024, 2048], respectively.
SpecGrad-5 and InferGrad-5 were also evaluated for comparison.

\begin{figure}[t]
\vspace{-1pt}
  \centering
  \includegraphics[width=\linewidth,clip]{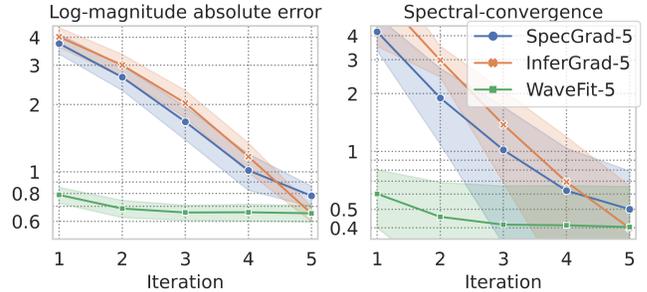} 
  \vspace{-20pt}
  \caption{Log-magnitude absolute error (left) and spectral convergence (right) of all intermediate outputs of SpecGrad-5, InferGrad-5, and WaveFit-5.
  Lower is better for both metrics.
  Solid lines mean the average, and colored area denote the standard-deviation. The scale of y-axis is log-scale.
  }
  \label{fig:obj_eval}
  \ReduceSpaceUnderFigure
\end{figure}

Figure~\ref{fig:obj_eval} shows the experimental results.
We can see that (i) both metrics for WaveFit decay on each iteration, and (ii) WaveFit outputs almost converge at three iterations.
In both metrics, WaveFit-5 was better than SpecGrad-5.
Although the objective scores of WaveFit-5 and InferGrad-5 were almost the same, we found the outputs of InferGrad-5 included small but noticeable artifacts.
A possible reason for these artifacts is that the DDPM-based models with a few iterations at inference needs to perform a large-level noise reduction at each iteration. This doesn't satisfy the small $\beta_t$ assumption of DDPM, which is required to make $q(\bm{x}_{t-1} \mid \bm{x}_t, \bm{c})$ Gaussian~\cite{sohl_2015,Ho_2020}. In contrast, WaveFit denoises an intermediate signal without adding random noise. Therefore, noise reduction level at each iteration can be small. This characteristics allow WaveFit to achieve higher audio quality with less iterations.
We provide intermediate output examples of these models in our demo page.\footnoteref{footnote:demo}

\ReduceSpaceAboveSubsection
\subsection{Comparison with WaveRNN and DDPM-based models}
\label{sec:experiment_results}

\begin{table}[ttt]
\caption{Real time factors (RTFs) and MOSs with their 95\% confidence intervals. Ground-truth means human natural speech.}
\label{tab:mos_result}
\centering
\begin{tabular}{ c | c c c }
\toprule
\textbf{Method} & \textbf{MOS} ($\uparrow$) & \textbf{RTF} ($\downarrow$) \\
\midrule
InferGrad-2 &  $3.68 \pm 0.07$  &  $0.030 \pm 0.00008$ \\	
WaveFit-2                        &  $\bm{4.13} \pm \bm{0.67}$  &  $\bm{0.028} \pm 0.0001$ \\	
\midrule
SpecGrad-3      &  $3.36 \pm 0.08$ &  $0.046 \pm 0.0018$ \\	
InferGrad-3 &  $4.03 \pm 0.07$ &  $0.045 \pm 0.0004$ \\	
WaveFit-3                        &  $\bm{4.33} \pm \bm{0.06}$ &  $\bm{0.041} \pm 0.0001$ \\	
\midrule
InferGrad-5  & $4.37 \pm 0.06$&  $0.072 \pm 0.0001$ \\
WaveRNN      & $4.41 \pm 0.05$&  $17.3 \pm 0.495$ \\	
WaveFit-5              &  $\bm{4.44} \pm \bm{0.05}$ &  $\bm{0.070} \pm 0.0020$ \\	
\midrule
Ground-truth & $4.50 \pm 0.05$  & $-$ \\
\bottomrule
\end{tabular}
\ReduceSpaceUnderTable
\end{table}

\begin{table}[ttt]
\vspace{-4pt}
\caption{Side-by-side test results with their 95\% confidence intervals. A positive score indicates that Method-A was preferred.}
\label{tab:sxs_result}
\centering
\begin{tabular}{ c c | c c }
\toprule
\textbf{Method-A} & \textbf{Method-B} & \textbf{SxS} & $\bm{p}$\textbf{-value}\\	
\midrule
WaveFit-3 & InferGrad-3  & $0.375 \pm 0.073$ & $0.0000$\\
WaveFit-3 & WaveRNN & $-0.051 \pm 0.044$ & $0.0027$\\	
\midrule
WaveFit-5 & InferGrad-5 & $0.063 \pm 0.050$ & $0.0012$\\
WaveFit-5 & WaveRNN & $-0.018 \pm 0.044$ & $0.2924$\\
WaveFit-5 & Ground-truth & $-0.027 \pm 0.037$ & $0.0568$\\
\bottomrule
\end{tabular}
\ReduceSpaceUnderTable
\end{table}

The MOS and RTF results and the SxS results using the 1,000 evaluation samples are shown in Tables~\ref{tab:mos_result} and \ref{tab:sxs_result}, respectively.
In all three iteration models, WaveFit produced better quality than both SpecGrad~\cite{specgrad} and InferGrad~\cite{InferGrad2022}.
As InferGrad used the same network architecture and adversarial loss as WaveFit, the main differences between them are (i) whether to compute loss value for all intermediate outputs, and (ii) whether to add random noise at each iteration.
These MOS and SxS results indicate that fixed-point iteration is a better strategy than DDPM for iterative-style neural vocoders.

InferGrad-3 was significantly better than SpecGrad-3.
The difference between InferGrad-3 and SpecGrad-3 is the use of $\lambda_{\mbox{\tiny IF}}$ only.
This result suggests that the hypothesis in Sec.~\ref{sec:ddpm_and_wavefit}, the mapping in DDPM-based neural vocoders only focuses on removing random components in input signals rather than moving the input signals towards the target, is supported.
Therefore, incorporating the difference between generated and target speech into the loss function of iterative-style neural vocoders is a promising approach.

On the RTF comparison, WaveFit models were slightly faster than SpecGrad and InferGrad models with the same number of iterations.
This is because DDPM-based models need to sample a noise waveform at each iteration, whereas WaveFit requires it only at the first iteration.

Although WaveFit-3 was worse than WaveRNN, WaveFit-5 achieved the naturalness comparable to WaveRNN and human natural speech; there were no significant differences in the SxS tests with $\alpha = 0.01$.
We would like to highlight that (i) the inference RTF of WaveFit-5 was over 240 times faster than that of WaveRNN, and (ii) the naturalness of WaveFit with 5 iterations is comparable to WaveRNN and human natural speech, whereas early DDPM-based models~\cite{wavegrad,diffwave} required over 100 iterations to achieve such quality.

\ReduceSpaceAboveSubsection
\subsection{Comparison with GAN-based models}
\label{sec:experiment_gan}

The MOS and SxS results using the LibriTTS 1,000 evaluation samples are shown in Table~\ref{tab:gan_result}.
These results show that WaveFit-5 was significantly better than MB-MelGAN, and there was no significant difference in naturalness between WaveFit-5 and HiFi-GAN $V$1.
In terms of the model complexity, the RTF and model size of WaveFit-5 are 0.07 and 13.8M, respectively, which are comparable to those of HiFi-GAN $V$1 reported in the original paper~\cite{Kong_2020}, 0.065 and 13.92M, respectively.
These results indicate that WaveFit-5 is comparable in the model complexity and naturalness with the well-tuned HiFi-GAN $V$1 model on LibriTTS dataset.

We found that some outputs from WaveFit-5 were contaminated by pulsive artifacts.
When we trained WaveFit using clean dataset recorded in an anechoic chamber (dataset used in the experiments of Sec.~\ref{sec:experiment_results}), such artifacts were not observed.
In contrast, the target waveform used in this experiment was not totally clean but contained some noise, which resulted in the erroneous output samples.
This result indicates that WaveFit models might not be robust against noise and reverberation in the training dataset.
We used the SpecGrad architecture from \cite{specgrad} both for WaveFit and DDPM-based models because we considered that the DDPM-based models are direct competitor of WaveFit and that using the same architecture provides a fair comparison.
After we realized the superiority of WaveFit over the other DDPM-based models, we performed comparison with GAN-based models, and hence the model architecture of WaveFit in the current version is not so sophisticated compared to GAN-based models.
Indeed, WaveFit-1 is significantly worse than GAN-based models, which can be heard form our audio demo.\footnoteref{footnote:demo}
There is a lot of room for improvement in the performance and robustness of WaveFit by seeking a proper architecture for $\mathcal{F}$, which is left as a future work.

\begin{table}[ttt]
\caption{Results of MOS and SxS tests on the LibriTTS dataset with their 95\% confidence intervals.  A positive SxS score indicates that WaveFit-5 was preferred.}
\label{tab:gan_result}
\centering
\begin{tabular}{ c | c | c c }
\toprule
\textbf{Method} & \textbf{MOS} ($\uparrow$) & \textbf{SxS}& $\bm{p}$\textbf{-value}\\
\midrule
MB-MelGAN     & $3.37 \pm 0.085$ &  $0.619 \pm 0.087$ &  $0.0000$ \\
HiFi-GAN $V$1 & $4.03 \pm 0.070$ &  $0.023 \pm 0.057$ &  $0.2995$ \\
Ground-truth  & $4.18 \pm 0.067$ &  $-0.089 \pm 0.052$ &  $0.0000$ \\
\midrule
WaveFit-5     & $3.98 \pm 0.072$ & $-$ & $-$ \\
\bottomrule
\end{tabular}
\ReduceSpaceUnderTable
\end{table}

% -------------------------------------------------------------------------
\section{Conclusion}
\label{sec:conclusion}

This paper proposed \textit{WaveFit}, which integrates the essence of GANs into a DDPM-like iterative framework based on fixed-point iteration.
WaveFit iteratively denoises an input signal like DDPMs while not adding random noise at each iteration.
This strategy was realized by training a DNN using a loss inspired by the concept of the fixed-point theory.
The subjective listening experiments showed that WaveFit can generate a speech waveform whose quality is better than conventional DDPM models. 
We also showed that the quality achieved by WaveFit with five iterations was comparable to WaveRNN and human natural speech, while
its inference speed was more than 240 times faster than WaveRNN.

% -------------------------------------------------------------------------
% Either list references using the bibliography style file IEEEtran.bst
\clearpage 
\balance
\bibliographystyle{IEEEtran}
%\footnotesize{
\bibliography{refs22}

\begin{thebibliography}{10}
\providecommand{\url}[1]{#1}
\def\UrlFont{\rmfamily}
\providecommand{\newblock}{\relax}
\providecommand{\bibinfo}[2]{#2}
\providecommand\BIBentrySTDinterwordspacing{\spaceskip=0pt\relax}
\providecommand\BIBentryALTinterwordstretchfactor{4}
\providecommand\BIBentryALTinterwordspacing{\spaceskip=\fontdimen2\font plus
\BIBentryALTinterwordstretchfactor\fontdimen3\font minus
  \fontdimen4\font\relax}
\providecommand\BIBforeignlanguage[2]{{%
\expandafter\ifx\csname l@#1\endcsname\relax
\typeout{** WARNING: IEEEtran.bst: No hyphenation pattern has been}%
\typeout{** loaded for the language `#1'. Using the pattern for}%
\typeout{** the default language instead.}%
\else
\language=\csname l@#1\endcsname
\fi
#2}}

\bibitem{sample_rnn}
S.~{Mehri}, K.~{Kumar}, I.~{Gulrajani}, R.~{Kumar}, S.~{Jain}, and J.~{Sotelo},
  ``{SampleRNN}: An unconditional end-to-end neural audio,'' in \emph{Proc.
  ICLR}, 2018.

\bibitem{tamamori2017speaker}
A.~Tamamori, T.~Hayashi, K.~Kobayashi, K.~Takeda, and T.~Toda,
  ``Speaker-dependent {WaveNet} vocoder.'' in \emph{Proc. Interspeech}, 2017.

\bibitem{waveglow}
R.~{Prenger}, R.~{Valle}, and B.~{Catanzaro}, ``{WaveGlow}: A flow-based
  generative network for speech synthesis,'' in \emph{Proc. ICASSP}, 2019.

\bibitem{waveflow}
W.~{Ping}, K.~{Peng}, K.~{Zhao}, and Z.~{Song}, ``{WaveFlow}: A compact
  flow-based model for raw audio,'' in \emph{Proc. ICML}, 2020.

\bibitem{tacotron2}
J.~{Shen}, R.~{Pang}, R.~J. {Weiss}, M.~{Schuster}, N.~{Jaitly}, Z.~{Yang},
  Z.~{Chen}, Y.~{Zhang}, Y.~{Wang}, R.~{Skerrv-Ryan}, R.~A. {Saurous},
  Y.~{Agiomvrgiannakis}, and Y.~{Wu}, ``Natural {TTS} synthesis by conditioning
  {WaveNet} on mel spectrogram predictions,'' in \emph{Proc. ICASSP}, 2018.

\bibitem{nat}
J.~Shen, Y.~Jia, M.~Chrzanowski, Y.~Zhang, I.~Elias, H.~Zen, and Y.~Wu,
  ``Non-attentive {Tacotron}: Robust and controllable neural {TTS} synthesis
  including unsupervised duration modeling,'' \emph{arXiv:2010.04301}, 2020.

\bibitem{parallel_tacotron}
I.~Elias, H.~Zen, J.~Shen, Y.~Zhang, Y.~Jia, R.~J. Weiss, and Y.~Wu, ``Parallel
  {Tacotron}: Non-autoregressive and controllable {TTS},'' in \emph{Proc.
  ICASSP}, 2021.

\bibitem{pngbert}
Y.~Jia, H.~Zen, J.~Shen, Y.~Zhang, and Y.~Wu, ``{PnG} {BERT}: Augmented {BERT}
  on phonemes and graphemes for neural {TTS},'' in \emph{Proc. Interspeech},
  2021.

\bibitem{fastspeech}
Y.~Ren, Y.~Ruan, X.~Tan, T.~Qin, S.~Zhao, Z.~Zhao, and T.-Y. Liu,
  ``{FastSpeech}: Fast, robust and controllable text to speech,'' in
  \emph{Proc. NeurIPS}, 2019.

\bibitem{fastspeech2}
Y.~Ren, C.~Hu, X.~Tan, T.~Qin, S.~Zhao, Z.~Zhao, and T.-Y. Liu, ``{FastSpeech}
  2: Fast and high-quality end-to-end text to speech,'' in \emph{Proc. Int.
  Conf. Learn. Represent. (ICLR)}, 2021.

\bibitem{vc_overview}
B.~Sisman, J.~Yamagishi, S.~King, and H.~Li, ``An overview of voice conversion
  and its challenges: From statistical modeling to deep learning,''
  \emph{IEEE/ACM Trans. Audio Speech Lang. Process.}, 2021.

\bibitem{wenchin_vc_2022}
W.-C. Huang, S.-W. Yang, T.~Hayashi, and T.~Toda, ``A comparative study of
  self-supervised speech representation based voice conversion,'' \emph{EEE J.
  Sel. Top. Signal Process.}, 2022.

\bibitem{translatotron}
Y.~Jia, R.~J. Weiss, F.~Biadsy, W.~Macherey, M.~Johnson, Z.~Chen, and Y.~Wu,
  ``Direct speech-to-speech translation with a sequence-to-sequence model,'' in
  \emph{Proc. Interspeech}, 2019.

\bibitem{translatotron2}
Y.~Jia, M.~T. Ramanovich, T.~Remez, and R.~Pomerantz, ``Translatotron 2:
  High-quality direct speech-to-speech translation with voice preservation,''
  in \emph{Proc. ICML}, 2022.

\bibitem{lee-etal-2022-direct}
A.~Lee, P.-J. Chen, C.~Wang, J.~Gu, S.~Popuri, X.~Ma, A.~Polyak, Y.~Adi, Q.~He,
  Y.~Tang, J.~Pino, and W.-N. Hsu, ``Direct speech-to-speech translation with
  discrete units,'' in \emph{Proc. 60th Annu. Meet. Assoc. Comput. Linguist.
  (Vol. 1: Long Pap.)}, 2022.

\bibitem{Maiti_waspaa_2019}
S.~{Maiti} and M.~I. {Mandel}, ``Parametric resynthesis with neural vocoders,''
  in \emph{Proc. IEEE WASPAA}, 2019.

\bibitem{Maiti_icassp_2020}
------, ``Speaker independence of neural vocoders and their effect on
  parametric resynthesis speech enhancement,'' in \emph{Proc. ICASSP}, 2020.

\bibitem{Su_2020}
J.~{Su}, Z.~{Jin}, and A.~{Finkelstein}, ``{HiFi-GAN}: High-fidelity denoising
  and dereverberation based on speech deep features in adversarial networks,''
  in \emph{Proc. Interspeech}, 2020.

\bibitem{Su_2021}
------, ``{HiFi-GAN-2}: Studio-quality speech enhancement via generative
  adversarial networks conditioned on acoustic features,'' in \emph{Proc. IEEE
  WASPAA}, 2021.

\bibitem{voice_filxer}
H.~{Liu}, Q.~{Kong}, Q.~{Tian}, Y.~{Zhao}, D.~L. {Wang}, C.~{Huang}, and
  Y.~{Wang}, ``{VoiceFixer}: Toward general speech restoration with neural
  vocoder,'' \emph{arXiv:2109.13731}, 2021.

\bibitem{saeki2021_IS}
T.~Saeki, S.~Takamichi, T.~Nakamura, N.~Tanji, and H.~Saruwatari,
  ``{SelfRemaster}: Self-supervised speech restoration with
  analysis-by-synthesis approach using channel modeling,'' in \emph{Proc.
  Interspeech}, 2022.

\bibitem{wavenet_codec}
W.~B. Kleijn, F.~S.~C. Lim, A.~Luebs, J.~Skoglund, F.~Stimberg, Q.~Wang, and
  T.~C. Walters, ``{WaveNet} based low rate speech coding,'' in \emph{Proc.
  ICASSP}, 2018.

\bibitem{wavenet_lossless_codec}
T.~Yoshimura, K.~Hashimoto, K.~Oura, Y.~Nankaku, and K.~Tokuda,
  ``{WaveNet}-based zero-delay lossless speech coding,'' in \emph{Proc. SLT},
  2018.

\bibitem{lpcnet_codec}
J.-M. Valin and J.~Skoglund, ``A real-time wideband neural vocoder at 1.6kb/s
  using {LPCNet},'' in \emph{Proc. Interspeech}, 2019.

\bibitem{sound_stream_codec}
N.~Zeghidour, A.~Luebs, A.~Omran, J.~Skoglund, and M.~Tagliasacchi,
  ``{SoundStream}: An end-to-end neural audio codec,'' \emph{IEEE/ACM Trans.
  Audio, Speech and Lang. Proc.}, 2022.

\bibitem{wavenet}
A.~{van den Oord}, S.~{Dieleman}, H.~{Zen}, K.~{Simonyan}, O.~{Vinyals},
  A.~{Graves}, N.~{Kalchbrenner}, A.~{Senior}, and K.~{Kavukcuoglu},
  ``{WaveNet}: A generative model for raw audio,'' \emph{arXiv:1609.03499},
  2016.

\bibitem{wavernn}
N.~{Kalchbrenner}, W.~{Elsen}, K.~{Simonyan}, S.~{Noury}, N.~{Casagrande},
  W.~{Lockhart}, F.~{Stimberg}, A.~{van den Oord}, S.~{Dieleman}, and
  K.~{Kavukcuoglu}, ``Efficient neural audio synthesis,'' in \emph{Proc. ICML},
  2018.

\bibitem{lpcnet}
J.-M. Valin and J.~Skoglund, ``{LPCNet}: Improving neural speech synthesis
  through linear prediction,'' in \emph{Proc. ICASSP}, 2019.

\bibitem{ParallelWaveNet}
A.~van~den Oord, Y.~Li, I.~Babuschkin, K.~Simonyan, O.~Vinyals, K.~Kavukcuoglu,
  G.~van~den Driessche, E.~Lockhart, L.~C. Cobo, F.~Stimberg, N.~Casagrande,
  D.~Grewe, S.~Noury, S.~Dieleman, E.~Elsen, N.~Kalchbrenner, H.~Zen,
  A.~Graves, H.~King, T.~Walters, D.~Belov, and D.~Hassabis, ``Parallel
  {W}ave{N}et: Fast high-fidelity speech synthesis.'' in \emph{Proc. ICML},
  2018.

\bibitem{normalizing_flow}
D.~J. Rezende and S.~Mohamed, ``Variational inference with normalizing flows,''
  in \emph{Proc. ICML}, 2015.

\bibitem{gan_goodfellow2014}
I.~Goodfellow, J.~Pouget-Abadie, M.~Mirza, B.~Xu, D.~Warde-Farley, S.~Ozair,
  A.~Courville, and Y.~Bengio, ``Generative adversarial nets,'' in \emph{Proc.
  NeurIPS}, 2014.

\bibitem{Donahue_2019}
C.~{Donahue}, J.~{McAuley}, and M.~{Puckette}, ``Adversarial audio synthesis,''
  in \emph{Proc. ICLR}, 2019.

\bibitem{Kong_2020}
J.~{Kong}, J.~{Kim}, and J.~Bae, ``{HiFi-GAN}: Generative adversarial networks
  for efficient and high fidelity speech synthesis,'' in \emph{Proc. NeurIPS},
  2020.

\bibitem{melgan}
K.~{Kumar}, R.~{Kumar}, T.~de~{Boissiere}, L.~{Gestin}, W.~Z. {Teoh},
  J.~{Sotelo}, A.~de~{Br\'{e}bisson}, Y.~{Bengio}, and A.~C. {Courville},
  ``{MelGAN}: Generative adversarial networks for conditional waveform
  synthesis,'' in \emph{Proc. Adv. Neural Inf. Process. Syst. (NeurIPS)}, 2019.

\bibitem{parallel_wavegan}
R.~Yamamoto, E.~Song, and J.-M. Kim, ``Parallel {WaveGAN}: A fast waveform
  generation model based on generative adversarial networks with
  multi-resolution spectrogram,'' in \emph{Proc. ICASSP}, 2020.

\bibitem{Yang_2021}
G.~{Yang}, S.~{Yang}, K.~{Liu}, P.~{Fang}, W.~{Chen}, and L.~{Xie},
  ``{Multi-band MelGAN}: Faster waveform generation for high-quality
  text-to-speech,'' in \emph{Proc. IEEE SLT}, 2021.

\bibitem{gan_vocoder}
J.~You, D.~Kim, G.~Nam, G.~Hwang, and G.~Chae, ``{GAN} vocoder:
  {M}ulti-resolution discriminator is all you need,'' \emph{arXiv:2103.05236},
  2021.

\bibitem{univnet}
W.~Jang, D.~Lim, J.~Yoon, B.~Kim, and J.~Kim, ``{UnivNet}: A neural vocoder
  with multi-resolution spectrogram discriminators for high-fidelity waveform
  generation,'' in \emph{Proc. Interspeech}, 2021.

\bibitem{kaneko2022istftnet}
T.~Kaneko, K.~Tanaka, H.~Kameoka, and S.~Seki, ``{iSTFTNet}: Fast and
  lightweight mel-spectrogram vocoder incorporating inverse short-time
  {Fourier} transform,'' in \emph{Proc. ICASSP}, 2022.

\bibitem{avocado_vocoder}
T.~Bak, J.~Lee, H.~Bae, J.~Yang, J.-S. Bae, and Y.-S. Joo, ``Avocodo:
  {G}enerative adversarial network for artifact-free vocoder,''
  \emph{arXiv:2206.13404}, 2022.

\bibitem{bigvgan}
S.-g. Lee, W.~Ping, B.~Ginsburg, B.~Catanzaro, and S.~Yoon, ``{BigVGAN}: A
  universal neural vocoder with large-scale training,''
  \emph{arXiv:2206.04658}, 2022.

\bibitem{wavegrad}
N.~{Chen}, Y.~{Zhang}, H.~{Zen}, R.~J. {Weiss}, M.~{Norouzi}, and W.~Chan,
  ``{WaveGrad}: Estimating gradients for waveform generation,'' in \emph{Proc.
  ICLR}, 2021.

\bibitem{diffwave}
Z.~{Kong}, W.~{Ping}, J.~{Huang}, K.~{Zhao}, and B.~{Catanzaro}, ``{DiffWave}:
  A versatile diffusion model for audio synthesis,'' in \emph{Proc. ICLR},
  2021.

\bibitem{Bddm022}
M.~W.~Y. {Lam}, J.~{Wang}, D.~{Su}, and D.~{Yu}, ``{BDDM}: Bilateral denoising
  diffusion models for fast and high-quality speech synthesis,'' in \emph{Proc.
  ICLR}, 2022.

\bibitem{priorgrad}
S.~{Lee}, H.~{Kim}, C.~{Shin}, X.~{Tan}, C.~{Liu}, Q.~{Meng}, T.~{Qin},
  W.~{Chen}, S.~{Yoon}, and T.-Y. {Liu}, ``{PriorGrad}: Improving conditional
  denoising diffusion models with data-dependent adaptive prior,'' in
  \emph{Proc. ICLR}, 2022.

\bibitem{specgrad}
Y.~Koizumi, H.~Zen, K.~Yatabe, N.~Chen, and M.~Bacchiani, ``{SpecGrad}:
  Diffusion probabilistic model based neural vocoder with adaptive noise
  spectral shaping,'' in \emph{Proc. Interspeech}, 2022.

\bibitem{okamoto2021}
T.~{Okamoto}, T.~{Toda}, Y.~{Shiga}, and H.~{Kawai}, ``Noise level limited
  sub-modeling for diffusion probabilistic vocoders,'' in \emph{Proc. ICASSP},
  2021.

\bibitem{Goel_2022}
K.~{Goel}, A.~{Gu}, C.~{Donahue}, and C.~R\'{e}, ``{It's Raw!} audio generation
  with state-space models,'' \emph{arXiv:2202.09729}, 2022.

\bibitem{InferGrad2022}
Z.~{Chen}, X.~{Tan}, K.~{Wang}, S.~{Pan}, D.~{Mandic}, L.~{He}, and S.~{Zhao},
  ``{InferGrad}: Improving diffusion models for vocoder by considering
  inference in training,'' in \emph{Proc. ICASSP}, 2022.

\bibitem{denoising_defusion_gans}
Z.~Xiao, K.~Kreis, and A.~Vahdat, ``Tackling the generative learning trilemma
  with denoising diffusion {GANs},'' in \emph{Proc. ICLR}, 2022.

\bibitem{diffgan_tts}
S.~Liu, D.~Su, and D.~Yu, ``{DiffGAN-TTS}: High-fidelity and efficient
  text-to-speech with denoising diffusion {GANs},'' \emph{arXiv:2201.11972},
  2022.

\bibitem{FixedPoint_DataSci}
P.~L. Combettes and J.-C. Pesquet, ``Fixed point strategies in data science,''
  \emph{IEEE Trans. Signal Process.}, 2021.

\bibitem{Ho_2020}
J.~{Ho}, A.~{Jain}, and P.~{Abbeel}, ``Denoising diffusion probabilistic
  models,'' in \emph{Proc. NeurIPS}, 2020.

\bibitem{demucs}
A.~Defossez, G.~Synnaeve, and Y.~Adi, ``Real time speech enhancement in the
  waveform domain,'' in \emph{Proc. Interspeech}, 2020.

\bibitem{unplugged2018}
G.~T. Buzzard, S.~H. Chan, S.~Sreehari, and C.~A. Bouman, ``Plug-and-play
  unplugged: Optimization-free reconstruction using consensus equilibrium,''
  \emph{SIAM J. Imaging Sci.}, 2018.

\bibitem{ICML_PnP2019}
E.~Ryu, J.~Liu, S.~Wang, X.~Chen, Z.~Wang, and W.~Yin, ``Plug-and-play methods
  provably converge with properly trained denoisers,'' in \emph{Proc. ICML},
  2019.

\bibitem{PesquetMMO2021}
J.-C. Pesquet, A.~Repetti, M.~Terris, and Y.~Wiaux, ``Learning maximally
  monotone operators for image recovery,'' \emph{SIAM J. Imaging Sci.}, 2021.

\bibitem{degli}
Y.~Masuyama, K.~Yatabe, Y.~Koizumi, Y.~Oikawa, and N.~Harada, ``Deep
  {Griffin}-{Lim} iteration,'' in \emph{Proc. ICASSP}, 2019.

\bibitem{degli_jstsp}
------, ``Deep {Griffin}–{Lim} iteration: Trainable iterative phase
  reconstruction using neural network,'' \emph{IEEE J. Sel. Top. Signal
  Process.}, 2021.

\bibitem{RED_PRO}
R.~Cohen, M.~Elad, and P.~Milanfar, ``Regularization by denoising via
  fixed-point projection {(RED-PRO)},'' \emph{SIAM J. Imaging Sci.}, 2021.

\bibitem{ConvexAnalysisBook}
H.~H. Bauschke and P.~L. Combettes, \emph{Convex Analysis and Monotone Operator
  Theory in Hilbert Spaces}.\hskip 1em plus 0.5em minus 0.4em\relax Springer,
  2017.

\bibitem{Moreau_Hybrid}
I.~Yamada, M.~Yukawa, and M.~Yamagishi, \emph{Minimizing the {Moreau} envelope
  of nonsmooth convex functions over the fixed point set of certain
  quasi-nonexpansive mappings}.\hskip 1em plus 0.5em minus 0.4em\relax
  Springer, 2011, pp. 345--390.

\bibitem{bond_prox}
N.~Parikh and S.~Boyd, ``Proximal algorithms,'' \emph{Found. Trends Optim.},
  2014.

\bibitem{seanet}
M.~Tagliasacchi, Y.~Li, K.~Misiunas, and D.~Roblek, ``{SEANet}: A multi-modal
  speech enhancement network,'' in \emph{Proc. Interspeech}, 2020.

\bibitem{SPMag_adaptLearn}
S.~Theodoridis, K.~Slavakis, and I.~Yamada, ``Adaptive learning in a world of
  projections,'' \emph{IEEE Signal Process. Mag.}, 2011.

\bibitem{kanbayashi}
T.~Hayashi, ``Parallel {WaveGAN} implementation with {Pytorch},''
  \url{github.com/kan-bayashi/ParallelWaveGAN}.

\bibitem{libritts}
H.~Zen, R.~Clark, R.~J. Weiss, V.~Dang, Y.~Jia, Y.~Wu, Y.~Zhang, and Z.~Chen,
  ``{LibriTTS}: A corpus derived from {LibriSpeech} for text-to-speech,'' in
  \emph{Proc. Interspeech}, 2019.

\bibitem{kingma2014adam}
D.~P. {Kingma} and J.~L. {Ba}, ``Adam: A method for stochastic optimization,''
  in \emph{Proc. ICLR}, 2015.

\bibitem{ged_loss}
A.~A. {Gritsenko}, T.~{Salimans}, R.~{van den Berg}, J.~{Snoek}, and
  N.~{Kalchbrenner}, ``A spectral energy distance for parallel speech
  synthesis,'' in \emph{Proc. NeurIPS}, 2020.

\bibitem{sohl_2015}
J.~Sohl-Dickstein, E.~Weiss, N.~Maheswaranathan, and S.~Ganguli, ``Deep
  unsupervised learning using nonequilibrium thermodynamic,'' in \emph{Proc.
  ICML}, 2015.

\end{thebibliography}
%}
\end{sloppy}
\end{document}